\documentclass{article}
\usepackage{chicago}
\usepackage{latexsym}
\usepackage{url}
\usepackage{graphicx}

\newcommand{\iec}{\mbox{i.\,e.\,}}

\newcommand{\egc}{\mbox{e.\,g.\,}}
\newcommand{\etc}{etc.\,\ }

\newcommand{\dr}[1]{\ensuremath{\mathrm{d} #1\,}}
\newcommand{\mc}[1]{\ensuremath{\mathcal{#1}}}

\newcommand{\ddt}{\ensuremath{\frac{\dr{}}{\dr{t}}}}

\newcommand{\pbp}[2]{\ensuremath{\frac{\partial #1}{\partial #2}}}

\newcommand{\ket}[1]{\ensuremath{\left|  #1 \right\rangle}}
\newcommand{\bra}[1]{\ensuremath{\left\langle #1 \right|}}
\newcommand{\bk}[2]{\ensuremath{\left\langle #1 | #2 \right\rangle}}
\newcommand{\proj}[2]{\ensuremath{\ket{#1} \bra{#2}}}
\newcommand{\tpk}[2]{\ensuremath{\ket{#1}\!\otimes\!\ket{#2}}}

\newcommand{\matel}[3]{\ensuremath{\bra{#1} #2 \ket{#3}}}

\newcommand{\denop}{\ensuremath{\rho}}

\newcommand{\be}{\begin{equation}}
\newcommand{\ee}{\end{equation}}
\begin{document}
\title{Decoherence and Ontology\linebreak (or: How I learned to stop worrying and love FAPP)}
\author{David Wallace}
\maketitle
\begin{quote}
The form of a philosophical theory, often enough, is: \emph{Let's try looking over here.}

\cite[p.\,31]{fodorrepresentations}
\end{quote}

\section{Introduction: taking physics seriously}\label{DW-intro}
NGC 1300 (shown in figure 1) is a spiral galaxy 65 million light years from Earth.\footnote{Source: \url{http://leda.univ-lyon1.fr/}. This photo taken from 
\url{http://hubblesite.org/gallery/album/galaxy/pr2005001a/}. [NB: issue of getting credit here.]}
 We have never been there, and (although I would love to be wrong about this) we will never go there; all we will ever know about NGC 1300 is what we can see of it from sixty-five million light years away, and what we can infer from our best physics.

\begin{figure}
\includegraphics[width=2.5in]{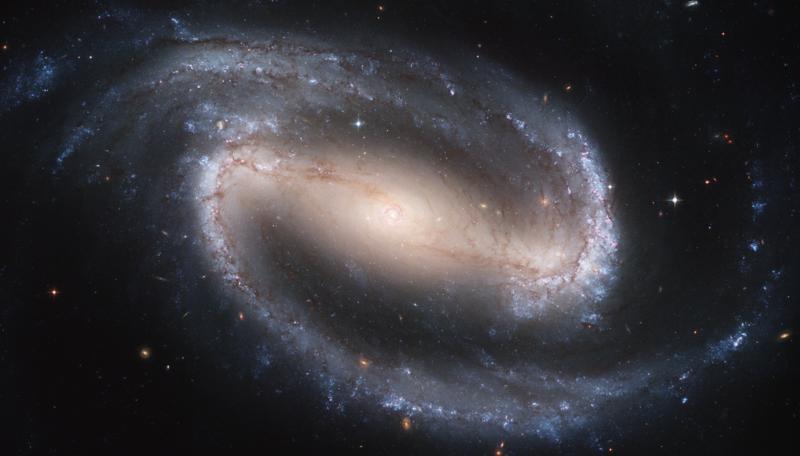}
\caption{The spiral galaxy NGC 1300}
\end{figure} 
Fortunately, ``what we can infer from our best physics'' is actually quite a lot. To take a particular example: our best theory of galaxies tells us that that hazy glow is actually made up of the light of hundreds of billions of stars; our best theories of planetary formation tell us that a sizable fraction of those stars have planets circling them, and our best theories of planetology tells us that some of those planets have atmospheres with such-and-such properties. And because I think that those ``best theories'' are actually pretty \emph{good} theories, I regard those inferences as fairly \emph{reliable}. That is: I think there actually \emph{are} atmospheres on the surfaces of some of the planets in NGC 1300, with pretty much the properties that our theories ascribe to them. That is: I think that those atmospheres \emph{exist}. I think that they are \emph{real}. I \emph{believe} in them. And I do so despite the fact that, at sixty-five million light years' distance, the chance of directly observing those atmospheres is nil.

I present this example for two reasons. The first is to try to demystify --- deflate, if you will --- the superficially ``philosophical'' --- even ``metaphysical'' --- talk that inevitably comes up in discussions of ``the ontology of the Everett interpretation''. Talk of ``existence'' and ``reality'' can sound too abstract to be relevant to physics (talk of ``belief'' starts to sound downright theological!) but in fact, when I say that ``I believe such-and-such is real'' I intend to mean no more than that it is on a par, evidentially speaking, with the planetary atmospheres of distant galaxies.

The other reason for this example brings me to the main claim of this paper. For the form of reasoning used above goes something like this: we have good grounds to take such-and-such physical theory seriously; such-and-such physical theory, taken literally, makes such-and-such ontological claim; therefore, such-and-such ontological claim is to be taken seriously.\footnote{Philosophers of science will recognise that, for reasons of space, and to avoid getting bogged down, I gloss over some subtle issues in the philosophy of science; the interested reader is invited to consult, \egc, \citeN{NSbook}, \citeN{psillosbook}, or \citeN{ladymanbook} for more on this topic.}

Now, if the mark of a serious scientific theory is its breadth of application, its explanatory power, its quantitative accuracy, and its ability to make novel predictions, then it is hard to think of a theory more ``worth taking seriously'' than quantum mechanics. So it seems entirely apposite to ask what ontological claims quantum mechanics makes, if taken literally, and to take those claims seriously in turn.

And quantum mechanics, taken literally, claims that we are living in a multiverse: that the world we observe around us is only one of countless quasi-classical universes (``branches'') all coexisting. In general, the other branches are no more observable than the atmospheres of NGC 1300's planets, but the theory claims that they exist, and so if the theory is worth taking seriously, we should take the branches seriously too. To belabour the point:

\begin{quote}
\textbf{According to our best current physics, branches are real.}
\end{quote}

 Everett was the first to recognise this, but for much of the ensuing fifty years it was overlooked: Everett's claim to be ``interpreting'' existing quantum mechanics, and de Witt's claim that ``the quantum formalism is capable of yielding its own interpretation'' were regarded as too simplistic, and much discussion on the Everett interpretation (even that produced by advocates such as \citeN{deutsch85}) took as read that the ``preferred basis problem'' --- the question of how the ``branches'' were to be defined --- could be solved only by adding something additional to the theory. Sometimes that ``something'' was additional physics, adding a multiplicity of worlds to the unitarily-evolving quantum state (\citeN{deutsch85,bellqmforcosmologists,barrettbook}). Sometimes it was a purpose-built theory of consciousness: the so-called ``many-minds theories'' (\citeN{lockwoodbook,albertloewerMM}). But whatever the details, the end result was a replacement of quantum mechanics by a new theory, and furthermore a new theory constructed specifically to solve the quantum measurement problem. No wonder interest in such theories was limited: if the measurement problem really does force us 
to change physics, hidden-variables theories like the de Broglie-Bohm theory\footnote{See\citeN{cushingbohmbook} and references therein for more information.}  or dynamical-collapse theories like the GRW theory\footnote{See \citeN{bassighirardireview} and references therein for more information.} seem to offer less extravagantly science-fictional options.

It now seems to be widely recognised that if Everett's idea really is worth taking seriously, it must be taken on Everett's own terms: as an understanding of what (unitary) quantum mechanics \emph{already} claims, not as a proposal for how to amend it. There is precedent for this: mathematically complex and conceptually subtle theories do not always wear their ontological claims on their sleeves. In general relativity, it took decades fully to understand that the existence of gravity waves and black holes really is a claim of the theory rather than some sort of mathematical artifact. 

Likewise in quantum physics, it has taken the rise of decoherence theory to illuuminate the structure of quantum physics in a way which makes the reality of the branches apparent. But twenty years of decoherence theory, together with the philosophical recognition that to be a ``world'' is not necessarily to be part of a theory's fundamental mathematical framework, now allow us to resolve --- or, if you like, to dissolve --- the preferred basis problem in a perfectly satisfactory way, as I shall attempt to show in the remainder of the paper.

\section{Emergence and Structure}\label{DW-emergence}

It is not difficult to see why Everett and de Witt's literalism seemed unviable for so long.
The axioms of unitary quantum mechanics say nothing of ``worlds'' or ``branches'': they speak only of a unitarily-evolving quantum state, and however suggestive it may be to write that state as a superposition of (what appear to be) classically definite states, we are not justified in speaking of those states as ``worlds'' unless they are somehow added into the formalism of quantum mechanics. As Adrian Kent put it in his influential \citeyear{kent} critique of Many-Worlds interpretations:
\begin{quote}
\ldots one can perhaps intuitively view the corresponding components [of the wave function] as describing a pair of independent worlds. But this intuitive interpretation goes beyond what the axioms justify: the axioms say nothing about the existence of multiple physical worlds corresponding to wave function components.
\end{quote} 
And so it appears that the Everettian has a dilemma: either the axioms of the theory must be modified to include explicit mention of ``multiple physical worlds'', or the existence of these multiple worlds must be some kind of illusion. But the dilemma is false. It is simply untrue that any entity not directly represented in the basic axioms of our theory is an illusion. Rather, science is replete with perfectly respectable entities which are nowhere to be found in the underlying microphysics. Douglas Hofstader and Daniel Dennett make this point very clearly:
\begin{quote}
Our world is filled with things that are neither mysterious and ghostly nor simply constructed out of the building blocks of physics. Do you believe in voices? How about haircuts? Are there such things? What are they? What, in the language of the physiicist, is a hole - not an exotic black hole, but just a hole in a piece of cheese, for instance? Is it a physical thing? What is a symphony? Where in space and time does ``The Star-Spangled Banner'' exist? Is it nothing but some ink trails in the Library of Congress? Destroy that paper and the anthem would still exist. Latin still \emph{exists} but it is no longer a living language. The language of the cavepeople of France no longer exists at all. The game of bridge is less than a hundred years old. What sort of a thing is it? It is not animal, vegetable, or mineral. 

These things are not physical objects with mass, or a chemical composition, but they are not purely abstract objects either - objects like the number pi, which is immutable and cannot be located in space and time. These things have birthplaces and histories. They can change, and things can happen to them. They can move about - much the way a species, a disease, or an epidemic can. We must not suppose that science teaches us that every \emph{thing} anyone would want to take seriously is identifiable as a collection of particles moving about in space and time.
\citeN[pp.\,6--7]{hofstadterdennett}
\end{quote}
The generic philosophy-of-science term for entities such as these is \emph{emergent}: they are not directly definable in the language of microphysics (try defining a haircut within the Standard Model!) but that does not mean that they are somehow independent of that underlying microphysics. To look in more detail at a particularly vivid example,\footnote{I first presented this example in \citeN{wallacestructure}.} consider Figure 2.\footnote{Photograph @ Philip Wallace, 2007. Reproduced with permission.}
\begin{figure}
\includegraphics[width=2.5in]{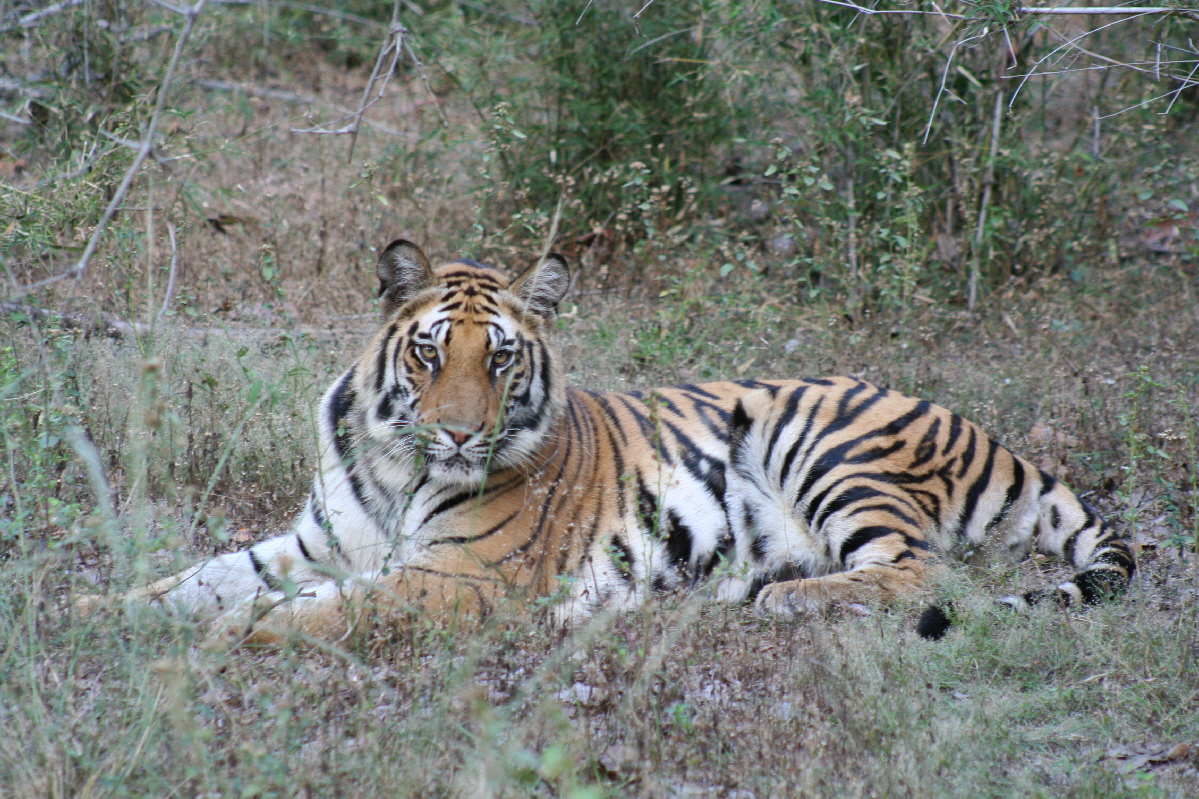}
\caption{An object not among the basic posits of the Standard Model}
\end{figure} Tigers are (I take it!) unquestionably real, objective physical objects, but the Standard model contains quarks, electrons and the like, but no tigers. Instead, tigers should be understood as patterns, or structures, \emph{within} the states of that microphysical theory.

To see how this works in practice, consider how we could go about
studying, say, tiger hunting patterns.  In principle --- and only in principle ---
the most reliable way to make predictions about these would be in terms of atoms and
electrons, applying molecular dynamics directly to the swirl of
molecules which make up, say, the Kanha National Park (one of the sadly diminishing places where Bengal tigers can be found).  In practice,
however (even ignoring the measurement problem itself!) this is clearly insane: no remotely imaginable computer would
be able to solve the $10^{35}$ or so simultaneous dynamical equations
which would be needed to predict what the tigers would do. 

Actually, the problem is even worse than this. For in a sense, we \emph{do} have  a computer capable of telling
us how the positions and momentums of all the molecules in the Kanha National Park change over time. It is called the Kanha National
Park. (And it runs in real time!)  Even if,  \emph{per impossibile}, we managed to build a computer simulation of the Park accurate down
to the last electron, it would tell us no more than what the Park itself tells us. It would provide no explanation of any of its complexity.
(It would, of course, be a superb vindication of our extant microphysics.)

If we want to understand the complex phenomena of the Park, and not just reproduce them, 
a more effective strategy can be found by studying the structures 
observable at the multi-trillion-molecule level of
description of this `swirl of molecules'.  At this level, we will
observe robust --- though not $100\%$ reliable --- regularities, which
will give us an alternative description of the tiger in a language of
cell membranes, organelles, and internal fluids.  The principles by which these 
interact will be derivable from the underlying microphysics, and will
involve various assumptions and approximations; hence very occasionally 
they will be found to fail.  Nonetheless, this slight riskiness in our
description is overwhelmingly worthwhile given the enormous gain in
usefulness of this new description: the language of  cell biology is
both explanatorily far more powerful, and practically far more useful,
than the language of physics for describing tiger behaviour.

Nonetheless it is still ludicrously hard work to study tigers in this
way.  To reach a really practical level of description, we again look
for patterns and regularities, this time in the behaviour of the cells
that make up individual tigers (and other living creatures which
interact with them).  In doing so we will reach yet another language,
that of zoology and evolutionary adaptationism, which describes the
system in terms of tigers, deer, grass, camouflage and so on.  This 
language is, of course, the norm in studying tiger hunting patterns, and
another (in practice very modest) increase in the riskiness of our
description is happily accepted in exchange for another phenomenal rise
in explanatory power and practical utility.

The moral of the story is:  there are structural facts about many microphysical systems
which, although perfectly real and objective (try telling a deer  that a nearby tiger is not objectively real) simply cannot be seen if we persist in describing those systems in purely microphysical language. Talk of zoology is of course grounded in cell biology, and cell biology in molecular physics, but the entities of zoology cannot be discarded in favour of the austere ontology of molecular physics alone. Rather, those entities are structures instantiated within the molecular physics, and the task of almost all science is to study structures of this kind. 

Of \emph{which} kind? (After all, ``structure'' and ``pattern'' are very broad terms: almost any arrangement of atoms might be regarded as some sort of pattern.) The tiger example suggests the following answer, which I have previously~\citeN[p.93]{wallacestructure} called ``Dennett's criterion'' in recognition of the very similar view proposed by Daniel Dennett \cite{dennettrealpatterns}:
\begin{quote}
\textbf{Dennett's criterion:} A macro-object is a pattern, and the existence of a pattern as a real thing depends on the usefulness --- in particular, the explanatory power and predictive reliability --- of theories which admit that pattern in their ontology. 
\end{quote}

Dennett's own favourite example is worth describing briefly in order to
show the ubiquity of this way of thinking: if I have a computer running
a chess program, I can in principle predict its next move from analysing
the electrical flow through its circuitry, but I have no
chance of doing this in practice, and anyway it will give me virtually
no understanding of that move.  I can achieve a vastly more effective
method of predictions if I know the program and am prepared to take the
(very small) risk that it is being correctly implemented by the
computer, but even this method will be practically very difficult to
use.  One more vast improvement can be gained if I don't concern myself
with the details of the program, but simply assume that whatever they
are, they cause the computer to play good chess.  Thus I move
successively from a language of electrons and silicon chips, through one
of program steps, to one of intentions, beliefs, plans and so forth ---
each time trading a small increase in risk for an enormous increase in
predictive and explanatory power.\footnote{It is, of course, highly
contentious to suppose that a chess-playing computer \emph{really}
believes, plans \etc  Dennett himself would embrace such claims (see
\citeN{dennettintentional} for an extensive discussion), but for the purposes of this section there
is no need to resolve the issue: the computer can be taken only to
`pseudo-plan', `pseudo-believe' and so on, without reducing the
explanatory importance of a description in such terms.}

Nor is this account restricted to the relation between physics and the rest of science: rather, it is ubiquitous within physics itself. 
Statistical mechanics provides perhaps the most important example of this: the temperature of bulk matter is an emergent property, salient because of its explanatory role in the behaviour of that matter. (It is a common error in textbooks to suppose that statistical-mechanical methods are used only because in practice we cannot calculate what each atom is doing separately: even if we could do so, we would be missing important, objective properties of the system in question if we abstained from statistical-mechanical talk.) But it is somewhat unusual because (unlike the case of the tiger) the principles underlying statistical-mechanical claims are (relatively!) straightforwardly derivable from the underlying physics. 

For an example from physics which is closer to the cases already discussed, consider the case of quasi-particles in solid-state physics. As is well known, vibrations in a (quantum-mechanical) crystal, although they can in principle be described entirely in terms of the individual crystal atoms and their quantum entanglement with one another, are in practice overwhelmingly simpler to describe in terms of `phonons' --- collective excitations of the crystal which behave like `real' particles in most respects. And furthermore, this sort of thing is completely ubiquitous in solid-state physics, with different sorts of excitation described in terms of different sorts of ``quasi-particle'' --- crystal vibrations are described in terms of phonons; waves in the magnetisation direction of a ferromagnet are described in terms of magnons, collective waves in a plasma are described in terms of plasmons, etc.

Are quasi-particles real? They can be created and annihilated; they can be scattered off one another; they can be detected (by, for instance, scattering them off ``real'' particles like neutrons); sometimes we can even measure their time of flight; they play a crucial part in solid-state explanations. We have no more evidence than this that ``real'' particles exist, and so it seems absurd to deny that quasi-particles exist --- and yet, they consist only of a certain pattern within the constituents of the solid-state system in question.

When \emph{exactly} are quasi-particles present? The question has no precise answer. It is essential in a quasi-particle formulation of a solid-state problem that the quasi-particles decay only slowly relative to other relevant timescales (such as their time of flight) and when this criterion (and similar ones) are met then quasi-particles are definitely present. When the decay rate is much too high, the quasi-particles decay too rapidly to behave in any `particulate' way, and the description becomes useless explanatorily; hence, we conclude that no quasi-particles are present. It is clearly a mistake to ask \emph{exactly} when the decay time is short enough (2.54 $\times$ the interaction time?) for quasi-particles not to be present, but the somewhat blurred boundary between states where quasi-particles exist and states when they don't should not undermine the status of quasi-particles as real, any more than the absence of a precise boundary to a mountain undermines the existence of mountains.

One more point about emergence will be relevant in what follows. In a certain sense emergence is a bottom-up process: knowledge of all the microphysical facts about the tiger and its environment suffices to derive all the tiger-level facts (in principle, and given infinite computing power). But in another sense it is a top-down process: no \emph{algorithmic} process, applied to a complex system, will tell us what higher-level phenomena to look for in that system. What makes it true that (say) a given lump of organic matter has intentions and desires is not something derivable algorithmically from that lump's microscopic constituents; it is the fact that, when it occurs to us to try interpreting its behaviour in terms of beliefs and desires, that strategy turns out to be highly effective.

\section{Decoherence and quasiclassicality}\label{DW-decoherence}

We now return to quantum mechanics, and to the topic of decoherence. In this section I will briefly review decoherence theory, in a relatively simple context (that of non-relativistic particle mechanics) and in the environment-induced framework advocated by, \egc, \citeN{joosetal} and Zurek (\citeyearNP{zurek91,zurek01review}). (An alternative formalism --- the ``decoherent histories'' framework advocated by, \egc, \citeN{gellmannhartle} and \citeN{halliwellhydrodynamic} --- is presented in the Introduction to this volume and in Halliwell's contribution to this volume.)

The basic setup is probably familiar to most readers. We assume that the Hilbert space \mc{H} of the system we are interested in  is factorised into ``system'' and ``environment'' subsystems, with Hilbert spaces  $\mc{H}_S$ and $\mc{H}_E$ respectively --- 
\be \mc{H}=\mc{H}_S \otimes \mc{H}_E.\ee
Here, the ``environment'' might be a genuinely external environment (such as the atmosphere or the cosmic microwave background); equally, it might be an ``internal environment'', such as the microscopic degrees of freedom of a fluid. For decoherence to occur, there needs to be some basis $\{\ket{\alpha}\}$ of $\mc{H}_S$ such that the dynamics of the system-environment interaction give us
\be
\tpk{\alpha}{\psi}\longrightarrow \tpk{\alpha}{\psi;\alpha}
\ee
and
\be\label{DW-orthogonal}
\bk{\psi;\alpha}{\psi;\beta}\simeq \delta(\alpha-\beta).
\ee
on timescales much shorter than those on which the system itself evolves.
(Here I use $\alpha$ as a ``schematic label''. In the case of a discrete basis $\delta(\alpha-\beta)$ is a simple Kronecker delta; in the case of a continuous basis, such as a basis of wavepacket states, then (\ref{DW-orthogonal}) should be read as requiring $\bk{\alpha}{\beta}\simeq 0$ unless $\alpha \simeq \beta$.)
In other words, the environment effectively ``measures'' the state of the system and records it. (The orthogonality requirement can be glossed as ``record states are distinguishable'', or as ``record states are dynamically sufficiently different'', or as ``record states can themselves be measured''; all, mathematically, translate into a requirement of orthogonality). Furthermore, we require that this measurement happens quickly: quickly, that is, relative to other relevant dynamical timescales for the system. (I use ``decoherence timescale'' to refer to the characteristic timescale on which the environment measures the system.)

Decoherence has a number of well-known consequences. Probably the best-known is diagonalisation of the system's density operator. Of course, \emph{any} density operator is diagonal in some basis, but decoherence guarantees that the system density operator will rapidly become diagonal in the $\{\ket{\alpha}\}$ basis, independently of its initial state: any initially non-diagonalised state will rapidly have its non-diagonal elements decay away.

Diagonalisation is a synchronic result: a constraint on the system at all times (or at least, on all time-intervals of order the decoherence timescale). But the more important consequence of decoherence is diachronic, unfolding over a period of time much longer than the decoherence timescale. Namely: because the environment is constantly measuring the system in the $\{\ket{\alpha}\}$ basis, any interference between distinct terms in this basis will be washed away. This means that, in the presence of decoherence, the system's dynamics is \emph{quasi-classical} in an important sense. Specifically: if we want to know the expectation value of any measurement on the system at some future time, it suffices to know what it would be were the system prepared in each particular $\ket{\alpha}$ at the present time (that is, to start the system in the state $\tpk{\alpha}{\psi}$ (for some environment state $\ket{\psi}$ whose exact form is irrelevant within broad parameters) and evolve it forwards to the future time), and then take a weighted sum of the resultant values. Mathematically speaking, this is equivalent to treating the system as though it were in some definite but unknown $\ket{\alpha}$. 

Put mathematically: suppose that the superoperator $\mc{R}$ governs the evolution of density operators over some given time interval, so that if the system intially has density operator $\denop$ then it has density operator $\mc{R}(\denop)$ after that time interval. Then in the presence of decoherence,
\be
\mc{R}(\denop)=\int \dr{\alpha}\matel{\alpha}{\denop}{\alpha}\mc{R}(\proj{\alpha}{\alpha}).
\ee
(Again: this integral is meant schematically, and should be read as a sum or an integral as appropriate.)

And of course, quasi-classicality is rather special. The reason, in general, that the quantum state cannot \emph{straightforwardly} be regarded as a probabilistic description of a determinate underlying reality is precisely that interference effects prevent the dynamics being quasi-classical. In the presence of decoherence, however, those interference effects are washed away.

\section{The significance of decoherence}

It might then be thought --- perhaps, at one point, it was thought --- that decoherence alone suffices to solve the measurement problem. For if decoherence picks out a certain basis for a system, and furthermore has the consequence that the dynamics of that system are quasi-classical, then --- it might seem --- we can with impunity treat the system not just as \emph{quasi}-classical but straightforwardly as classical. In effect, this would be to use decoherence to give a precise and observer-independent definition of the collapse of the wavefunction: the quantum state evolves unitarily as long as superpositions which are not decohered from one another do not occur; when such superpositions do occur, the quantum state collapses instantaneously into one of them. To make this completely precise would require us to discretize the dynamics so that the system evolves in discrete time steps rather than continuously The decoherent-histories formalism mentioned earlier is a rather more natural mathematical arena to describe this than the continuous formalism I developed in section \ref{DW-decoherence}, but the result is the same in any case: decoherence allows us to extract from the unitary dynamics a space of \emph{histories} (strings of projectors onto decoherence-preferred states) and to assign probabilities to each history in a consistent way (\iec, without interference effects causing the probability calculus to be violated.

From a conceptual point of view there is something a bit odd about this strategy. Decoherence is a dynamical process by which two components of a complex entity (the quantum state) come to evolve independently of one another, and it occurs due to rather high-level, emergent consequences of the particular dynamics and initial state of our Universe. Using this rather complex high-level process as a criterion to define a new fundamental law of physics is, at best, an exotic variation of normal scientific practice. (To take a philosophical analogy, it would be as if psychologists constructed a complex theory of the brain, complete with a physical analysis of memory, perception, reasoning and the like --- and then decreed that, as a new fundamental law of physics (and not a mere definition), a system was conscious if and only if it had those physical features.\footnote{As it happens, this is not a straw man: David Chalmers has proposed something rather similar. See \citeN{chalmersbook} for an exposition, and \citeN{dennettfantasy} for some sharp criticism.})

Even aside from such conceptual worries, however, a pure-decoherence solution to the measurement problem turns out to be impossible on technical grounds: the decoherence criterion is both too strong, and too weak, to pick out an appropriate set of classical histories from the unitary quantum dynamics.

That decoherence is too \emph{strong} a condition should be clear from the language of section \ref{DW-decoherence}. Everything there was approximate, effective, for-all-practical-purposes: decoherence occurs on short timescales (not instantaneously); it causes interference effects to become negligible (not zero); it approximately diagonalises the density operator (not exactly); it approximately selects a preferred basis (not precisely).  And while approximate results are fine for calculational shortcuts or for 
emergent phenomena, they are most unwelcome when we are trying to define new fundamental laws of physics. (Put another way, a theory cannot be 99.99804\% conceptually coherent.)

That it is too \emph{weak} is more subtle, but ultimately even more problematic. There are simply \emph{far too many} bases picked out by decoherence --- in the language of section \ref{DW-decoherence} there are far too many system-environment splits which give rise to an approximately decoherent basis for the system; in the language of decoherent histories, there are far too many choices of history that lead to consistent classical probabilities. Worse, there are good reasons (cf~\citeN{dowkerkent})to think that many, many of these histories are wildly non-classical.

What can be done? Well, if we turn away from the abstract presentation of decoherence theory, and look at the concrete models (mathematical models and computer simulations) to which decoherence has been applied, and if, in those models, we make the sort of system/environment split that fits our natural notion of environment (so that we take the environment, as suggested previously, to be --- say --- the microwave background radiation, or the residual degrees of freedom of a fluid once its bulk degrees of freedom have been factored out), then we find two things. 

Firstly: The basis picked out by decoherence is approximately a coherent-state basis: that is, it is a basis of wave-packets approximately localised in both position and momentum.
And secondly: The dynamics is quasi-classical not just in the rather abstract, bloodless sense used in section \ref{DW-decoherence}, but in the sense that the behaviour of those wave-packets approximates the behaviour predicted by classical mechanics.

In more detail: let $\ket{q,p}$ denote a state of the system localised around phase-space point $(q,p)$. Then  decoherence ensures that the state of the system+environment at any time $t$ can be written as 
\be
\ket{\Psi}=\int\dr{q}\dr{p}\alpha{q,p;t}\tpk{q,p}{\epsilon(q,p)}
\ee
with $\bk{\epsilon(q,p)}{\epsilon(q',p')}=0$ unless $q\simeq q'$ and $p\simeq p'$. 
The conventional (\iec, textbook) interpretation of quantum mechanics tells us that $|\alpha(q,p)|^2$ is the probability density for finding the system in the vicinity of phase-space point $(q,p)$.\footnote{At a technical level, this requires the use of phase-space POVMs (\iec, positive operator valued measures, a generalisation of the standard projection-valued measures; see, \egc, \citeN{nielsenchuang} for details): for instance,  the continuous family $\{
N\proj{q,p}{q,p}\}$ is an appropriate POVM for suitably-chosen normalisation constant $N$. Of course, this or any phase-space POVM can only be defined for measurements of accuracy $\leq \hbar$.} Then in the presence of decoherence, $|\alpha|^2(q,p)$ evolves, to a good approximation, like a \emph{classical} probability density on phase space: it evolves, approximately, under the Poisson equations
\be\ddt\left(|\alpha(q,p)|^2\right)\simeq \pbp{H}{q}\pbp{|\alpha(q,p)|^2}{p}-\pbp{H}{p}\pbp{|\alpha(q,p)|^2}{q}
\ee
where $H(q,p)$ is the Hamiltonian.  

On the assumption that the system is classically non-chaotic (chaotic systems add a few subtleties), this is equivalent to the claim that each individual wave-packet follows a classical trajectory on phase space. Structurally speaking, the dynamical behaviour of each wave-packet is the same as the behaviour of a macroscopic classical system. And if there are multiple wave-packets, the system is dynamically isomorphic to a collection of independent classical systems.

(\emph{Caveat}: this does not mean that the wave-packets are actually evolving on phase space. 
If phase space is understood as the position-momentum space of a collection of classical point particles, then \emph{of course} the wave-packets are not evolving on phase space. They are evolving on a space isomorphic to phase space. Henceforth when I speak of phase space, I mean this space, not the ``real'' phase space.)

So: if we pick a particular choice of system-environment split, we find a ``strong'' form of quasi-classical behaviour: we find that the system is isomorphic to a collection of dynamically independent simulacra of a classical system. We did not find this isomorphism by some formal algorithm; we found it by making a fairly unprincipled choice of system-environment split and then noticing that that split led to interesting behaviour. The interesting behaviour is no less real for all that.

We can now see that all three of the objections at the start of this section point at the same --- fairly obvious --- fact: decoherence is an emergent process occurring \emph{within} an already-stated microphysics: unitary quantum mechanics. It is not a mechanism to define a part \emph{of} that microphysics. If we think of quasiclassical histories as emergent in this way, then
\begin{itemize}
\item The ``conceptual mystery'' dissolves: we are not using decoherence to define a dynamical collapse law, we are just using it as a (somewhat pragmatic) criterion for when quantum systems display quasiclassical behaviour.
\item There is nothing problematic about the approximateness of the decoherence process: as we saw in section \ref{DW-emergence}, this is absolutely standard features of emergence.
\item Similarly, the fact that we had no algorithmic process to tell us in a bottom-up way what system-environment splits would lead to the discovery of interesting structure is just a special case of section \ref{DW-emergence}'s observation that emergence is in general a somewhat top-down process.
\end{itemize}
Each decoherent history is an emergent structure within the underlying quantum state, on a par with tigers, tables, and the other emergent objects of section \ref{DW-emergence} --- that is, on a par with practically all of the objects of science, and no less real for it.

But the price we pay for this account is that, if the fundamental dynamics are unitary, at the fundamental level there is no collapse of the quantum state. There is just a dynamical process --- decoherence --- whereby certain components of that state become 
dynamically autonomous of one another. Put another way: if each decoherent history is an emergent structure within the underlying microphysics, and if the underlying microphysics doesn't do anything to prioritise one history over another (which it doesn't) then all the histories exist. That is: a unitary quantum theory with emergent, decoherence-defined quasi-classical histories is a many-worlds theory.

\section{Simulation or reality?}

At this point, a skeptic might object:
\begin{quote}
All you have shown is that certain features of the unitarily-evolving quantum state are isomorphic to a classical world. If that's true, the most it shows that the quantum state is running a simulation of the classical world. But I didn't want to recover a \emph{simulation} of the world. I wanted to recover \emph{the world}.
\end{quote} 

I rather hope that this objection is a straw man: as I attempted to illustrate in section \ref{DW-emergence}, this kind of structural story about higher-level ontology (the classical world is a structure instantiated in the quantum state) is totally ubiquitous in science. But it seems to be a common enough thought (at least in philosophical circles) to be worth engaging with in more detail.

Note firstly that the very assumption that a certain entity which is structurally like our world is not \emph{our world} is manifestly question-begging. How do we know that space is three-dimensional? We look around us. How do we know that we are seeing something fundamental rather than emergent? We don't; all of our observations (\emph{pace} Maudlin, this volume) are structural observations, and only the sort of aprioristic knowledge now fundamentally discredited in philosophy could tell us more.

Furthermore, physics itself has always been totally relaxed about this sort of possibility. A few examples will suffice:
\begin{itemize}
\item Solid matter --- described so well, and in such accord with our observations, in the language of continua --- long ago turned out to be only emergently continuous, only emergently solid. 
\item Just as solid state physics deals with emergent quasi-particles, so --- according to modern ``particle physics'' --- elementary particles themselves turn out to be emergent from an underlying quantum field. Indeed, the ``correct'' --- that is, most explanatorily and predictively useful --- way of dividing up the world into particles of different types turns out to depend on the energy scales at which we are working.\footnote{The best known example of this phenomenon occurs in quantum chromodynamics: treating the quark field in terms of approximately-free quarks works well at very high energies, but at lower energies the appropriate particle states are hadrons and mesons; see, \egc, \citeN{chengli} and references therein for details. For a more mathematically tractable example (in which even the correct choice of whether particles are fermionic or bosonic is energy-level-dependent), see chapter 5 of \citeN{coleman}, esp. pp.\,246--253.}
\item The idea that particles should be emergent from some field theory is scarcely new: in the 19th century there was much exploration of the idea that particles were topological structures within some classical continuum (cf \citeN{epple}), and later, \citeN{wheelergeometrodynamics} proposed that matter was actually just a structural property of a very complex underlying spacetime. Neither proposal eventually worked out, but for technical reasons: the proposals themselves were seen as perfectly reasonable. 
\item The various proposals to quantize gravity have always been perfectly happy with the idea that space itself would turn out to be emergent. From Borel dust to non-commutative geometry to spin foam, program after program has been happy to explore the possibility that spacetime is only emergently a four-dimensional continuum.\footnote{For the concept of Borel dust, see \citeN[p.1205]{mtw}; for references on non-commutative geometry, see http://www.alainconnes.org/en/downloads.php; for references on spin foam, see \citeN{rovelli}.}
\item String theory, currently the leading contender for a quantum theory of gravity, regards spacetime as fundamentally high-dimensional and only emergently four-dimensional, and the recent development of the theory makes the nature of that emergence more and more indirect (it has been suggested, for instance, that the ``extra'' dimensions may be several centimetres across\footnote{For a brief introduction to this proposal, see \citeN[chapter 29]{dinestring}.}). The criterion for emergence, here as elsewhere, are dynamical: if the functional integrals that define the cross-sections have the approximate functional form of functional integrals of fields on four-dimensional space, that is regarded as sufficient to establish emergence.
\end{itemize}

Leaving aside these sorts of naturalistic\footnote{I use ``naturalism'' in Quine's sense (\cite{quinenaturalism}): a naturalistic philosophy is one which regards our best science as the only good  guide to our best epistemology,} considerations, we might ask: \emph{what} distinguishes a simulation of a thing from the thing itself? It seems to me that there are two relevant distinctions:
\begin{enumerate}
\item[Dependency:]  Tigers don't interact with simulations of tigers; they interact with the computers that run those simulations. The simulations are instantiated in ``real'' things, and depend on them to remain in existence.
\item[Parochialism:] Real things have to be made of a certain sort of stuff, and/or come about in a certain sort of way. Remarkably tiger-like organisms in distant galaxies are not tigers; synthetic sparkling wine, however much it tastes like champagne, is not champagne unless its origins and makeup fit certain criteria.
\end{enumerate}
Now, these considerations are themselves problematic. (Is a simulation of a person themselves a person? --- see \cite{hofstadterturing} for more thoughts on these matters). But, as I hope is obvious, both considerations are question-begging in the context of the Everett interpretation: only if we begin with the assumption that our world is instantiated in a certain way can we argue that Everettian branches are instantiated in a relevantly different way.

\section{How many worlds?}

We are now in a position to answer one of the most commonly asked questions about the Everett interpretation,\footnote{Other than ``and you believe this stuff?!'', that is.} namely: how much branching actually happens? As we have seen, branching is caused by any process which magnifies microscopic superpositions up to the level where decoherence kicks in, and there are basically three such processes:
\begin{enumerate}
\item Deliberate human experiments: Schr\"{o}dinger's cat, the two-slit experiment, Geiger counters, and the like.
\item ``Natural quantum measurements'', such as occur when radiation causes cell mutation.
\item Classically chaotic processes, which cause small variations in initial conditions to grow exponentially, and so which cause quantum states which are initially spread over small regions in phase space to spread over macroscopically large ones. (See \citeN{zurek94} for more details; I give a conceptually oriented introduction in \citeN{wallacestatmech}.)
\end{enumerate}
The first is a relatively recent and rare phenomenon, but the other two are ubiquitous. Chaos, in particular, is everywhere, and where there is chaos, there is branching (the weather, for instance, is chaotic, so there will be different weather in different branches). Furthermore, there is no sense in which these phenomena lead to a naturally \emph{discrete} branching process. Quantum chaos gives rise to macroscopic superpositions, and so to decoherence and to the emergence of a branching structure, but that structure  has no natural ``grain''. To be sure, by choosing a certain discretisation of (phase-)space and time, a discrete branching structure will emerge, but a finer or coarser choice would also give branching. And there is no ``finest'' choice of branching structure: as we fine-grain our decoherent history space, we will eventually reach a point where interference between branches ceases to be negligible, but there is no precise point where this occurs. As such, the question ``how many branches are there?'' does not, ultimately, make sense.

This may seem paradoxical --- certainly, it is not the picture of ``parallel universes'' one obtains from science fiction. But as we have seen in  this chapter, it is commonplace in emergence for there to be some indeterminacy (recall: when \emph{exactly} are quasi-particles of a certain kind present?) And nothing prevents us from making statements like:
\begin{quote}
Tomorrow, the branches in which it is sunny will have combined weight 0.7
\end{quote}
--- the combined weight of all branches having a certain macroscopic property is very (albeit not precisely) well-defined. It is only if we ask: "\emph{how many} branches are there in which it is sunny", that we end up asking a question which has no answer.

This bears repeating, as it is central to some of the arguments about probability in the Everett interpretation:
\begin{quote}
Decoherence causes the Universe to develop an emergent branching structure. The existence of this branching is a robust (albeit emergent) feature of reality; so is the mod-squared amplitude for any \emph{macroscopically described} history. But there is \emph{no} non-arbitrary decomposition of macroscopically-described histories into ``finest grained'' histories, and \emph{no} non-arbitrary way of counting those histories. 
\end{quote}
(Or, put another way: asking how many worlds there are is like asking how many experiences you had yesterday, or how many regrets a repentant criminal has had. It makes perfect sense to say that you had many experiences or that he had many regrets; it makes perfect sense to list the most important categories of either; but it is a non-question to ask \emph{how many}.)

If this picture of the world seems unintuitive, a metaphor may help. 

\begin{enumerate}
\item Firstly, imagine a world consisting of a very thin, infinitely long and wide, slab of matter, in which various complex internal processes are occurring --- up to and including the presence of intelligent life, if you like.  In particular one might imagine various forces acting in the plane of the slab, between one part and another.
\item Now, imagine stacking many thousands of these slabs one atop the other, but without allowing them to interact at all. If this is a ``many-worlds theory'', it is a many-worlds theory only in the sense of the philosopher David Lewis \cite{lewisplurality}: none of the worlds are dynamically in contact, and no (putative) inhabitant of any world can gain empirical evidence about any other.
\item Now introduce a weak force normal to the plane of the slabs --- a force with an effective range of 2-3 slabs, perhaps, and a force which is usually very small compared to the intra-slab force. Then other slabs will be detectable from within a slab but will not normally have much effect on events within a slab. If this is a many-worlds theory, it is a science-fiction-style many-worlds theory (or maybe a Phillip Pullman or C.S. Lewis many-worlds theory\footnote{See, for instance, Pullman's \emph{Northern Lights} or Lewis's \emph{The Magician's Nephew}.}): there are many worlds, but each world has its own distinct identity.
\item Finally, turn up the interaction sharply: let it have an effective range of several thousand slabs, and let it be comparable in strength (over that range) with characteristic short-range interaction strengths within a slab. Now, dynamical processes will not be confined to a slab but will spread over hundreds of adjacent slabs; indeed, \emph{evolutionary} processes will not be confined to a slab, so living creatures in this universe will exist spread over many slabs. At this point, the boundary between slabs becomes epiphenomenal. Nonetheless, this theory is \emph{stratified} in an important sense: dynamics still occurs predominantly along the horizontal axis and events hundreds of thousands of slabs away from a given slab are dynamically irrelevant to that slab.\footnote{Obviously there would be ways of constructing the dynamics so that this was not the case: if signals could easily propagate vertically, for instance, the stratification would be lost. But it's  only a thought experiment, so we can construct the dynamics how we like.} One might well, in studying such a system, divide it into layers thick relative to the range of the inter-slab force --- and emergent dynamical processes in those layers would be no less real just because the exact choice of layering is arbitrary. 
\end{enumerate}
Ultimately, though, that a theory of the world is ``unintuitive'' is no argument against it, provided it can be cleanly described in mathematical language. Our intuitions about what is ``reasonable'' or ``imaginable'' were designed to aid our ancestors on the savannahs of Africa, and the Universe is not obliged to conform to them.

\section{Conclusion}

The claims of the Everett interpretation are:
\begin{itemize}
\item At the most fundamental level, the quantum state is all there is
–-- quantum mechanics is about the structure and evolution of the quantum state
in the same way that (e.g.) classical field theory is about the
structure and evolution of the fields. 
\item As such, the ``Everett interpretation of quantum mechanics'' is just quantum mechanics itself, taken literally (or, as a philosopher of science might put it, Realist-ically) as a description of the Universe. De Witt has been widely criticized for his claim that "the formalism of quantum mechanics yields its own interpretation'' \cite{dewitt}, but there is nothing mysterious or Pythagorean about it: \emph{every} scientific theory yields its own interpretation, or rather (cf David Deutsch's contribution to this volume) the idea that one can divorce a scientific theory from its interpretation is confused.
\item ``Worlds'' are mutually dynamically isolated structures instantiated within the
quantum state, which are structurally and dynamically
``quasiclassical''.
\item The existence of these ``worlds'' is established by decoherence theory.
\end{itemize}
No \emph{postulates} about the worlds have needed to be added: the question of whether decoherence theory does indeed lead to the emergence of a quasiclassical branching structure is (at least in principle) settled \emph{a priori} for any particular quantum theory once we know the initial state. It is not even a \emph{postulate} that decoherence is the source of all ``worlds''; indeed, certain specialised experiments --- notably, some algorithms on putative quantum computers --- would also give rise to multiple quasiclassical worlds at least locally; cf. \citeN{deutschfabric}.\footnote{Since much hyperbole and controversy surrounds claims about Everett and quantum computation, let me add two deflationary comments:
\begin{enumerate}
\item There is no particular reason to assume that \emph{all} or even \emph{most} interesting quantum algorithms operate by any sort of ``quantum parallelism'' (that is: by doing different classical calculations in a large number of terms in a superposition and then interfering them). Indeed, Grover's algorithm does not seem open to any such analysis. But Shor's algorithm, at least, does seem to operate in this way.
\item The correct claim to make about Shor's algorithm is not (\emph{pace} \cite{deutschfabric}) that the calculations \emph{could not} have been done other than by massive parallelism, but simply that the actual explanation of how they \emph{were} done --- that is, the workings of Shor's algorithm --- does involve massive parallelism.
\end{enumerate}
For some eloquent (albeit, in my view, mistaken) criticisms of the link between quantum computation and the Everett interpretation, see \citeN{steaneoneuniverse}.}

I will end this discussion on a lighter note, aimed at a slightly different audience. I have frequently talked to physicists who accept Everett's interpretation, accept (at least when pressed!) that this entails a vast multiplicity of quasiclassical realities, but reject the ``many-worlds'' label for the interpretation --- thhey prefer to say that there is only one world but it contains many non- or hardly-interacting quasiclassical parts.

But, as I hope I have shown, the ``many worlds'' of Everett's many-worlds interpretation are not fundamental additions to the theory. Rather, they are emergent entities which, according to the theory, are present in large numbers. In this sense, the Everett interpretation is a ``many-worlds theory'' in just the same sense as African zoology is a ``many-hippos theory'': that is, there are entities whose existence is entailed by the theory which deserve the name ``worlds''. So, to Everettians cautious about the ``many-worlds'' label, I say: come on in, the water's lovely.

\end{document}